\documentclass[epsf]{elsart}

\usepackage{graphicx}
\usepackage{amssymb}

\journal{Physica A}

\begin{document}
\begin{frontmatter}

\title{Michaelis-Menten Dynamics in Complex Heterogeneous Networks}

\author[CONDMAT,ICMA,BIFI]{J. G\'omez-Garde\~nes},
\author[BIFI]{Y. Moreno}\ead{yamir@unizar.es}  and
\author[CONDMAT,ICMA,BIFI]{L.M. Flor\'{\i}a}

\address[CONDMAT]{Dpt. de F\'{\i}sica de la Materia Condensada,
Universidad de Zaragoza, 50009 Zaragoza, Spain.}
\address[ICMA]{Dept. de Teor\'{\i}a y Simulaci\'on de Sistemas Complejos,
Instituto de Ciencia de Materiales de Arag\'on (ICMA), C.S.I.C.-Universidad
de Zaragoza, 50009 Zaragoza, Spain.}
\address[BIFI]{Instituto de Biocomputaci\'on y F\'{\i}sica de Sistemas
Complejos (BIFI), Universidad de Zaragoza, Zaragoza 50009, Spain}

\begin{abstract}
  
  Biological networks have been recently found to exhibit many topological
  properties of the so-called complex networks. It has been reported that they
  are, in general, both highly skewed and directed. In this paper, we report on
  the dynamics of a Michaelis-Menten like model when the topological features
  of the underlying network resemble those of real biological networks.
  Specifically, instead of using a random graph topology, we deal with a
  complex heterogeneous network characterized by a power-law degree
  distribution coupled to a continuous dynamics for each network's component.
  The dynamics of the model is very rich and stationary, periodic and chaotic
  states are observed upon variation of the model's parameters. We characterize
  these states numerically and report on several quantities such as the
  system's phase diagram and size distributions of clusters of stationary,
  periodic and chaotic nodes. The results are discussed in view of recent
  debate about the ubiquity of complex networks in nature and on the basis of
  several biological processes that can be well described by the dynamics
  studied.
\end{abstract}

\end{frontmatter}

\section{Introduction}
\label{s.intro}

The discovery that many seemingly diverse systems, both natural and
man-made, can be represented as networks with similar topological
properties has driven a great body of research work in the last few
years \cite{strogatz,book1,book2}. Network modeling has become a
useful and common tool in fields as diverse as communication
\cite{ref7,alexei,book3}, biological \cite{ref5,ref6} and social
systems \cite{pnas}. For instance, biological applications of network
modeling range from the design of new drugs \cite{batt} to a better
understanding of basic cellular processes \cite{ref5}. In
technological networks such as the Internet and the world-wide-web,
the challenges include the design of new communication strategies in
order to provide faster access time to millions of users \cite{pablo},
the implementation of better algorithms for database exchange and
information dissemination \cite{maziar,mmp04}, and the understanding of the
topological features with the final goal of protecting the networks
against random failures, intentional attacks and virus spreading
\cite{book3,newman00,pv01a,moreno02,mgp03,vm03}.

These networks are described by several characteristics. Among all
the properties that can be studied, one usually finds that real-world
networks are small-worlds (SW), which means that the average distance
between two arbitrarily chosen nodes scales with the system size only
logarithmically \cite{book1,book2}. Besides, the structural complexity
of these networks is characterized by the number of interacting
partners (connectivity $k$) of a given element (node). Surprisingly,
the majority of real-world networks studied so far display a
distribution of links that follows a power-law $P(k)\sim k^{-\gamma}$
(termed scale-free networks), with $\gamma\leq 3$
\cite{strogatz,book1}.

Biological networks at all levels of organization are nowadays the subject of
intense experimental and theoretical research. Recent analysis of
protein-protein interaction networks has provided new useful insights into
biological essentiality at this level of organization \cite{modularity}. It is
also believed that a better comprehension of gene and protein networks will
help to elucidate the functions of a large fraction of proteins whose functions
are unknown \cite{avbiot}. Moreover, it is a major challenge the discovery of
how biological entities interact to perform specific biological processes and
tasks, as well as how their functioning is so robust under variations of
internal and external parameters \cite{theo1,exp1,exp2}. 

It has been recently shown \cite{exp3} that regulatory genes interact forming a
complex interconnected network. This network is both {\em directed} and {\em
  highly skewed} for the yeast {\em Saccharomyces cerevisiae}. This means that
there are a few regulatory genes that interact with many others but most of the
genes only participate in a few processes. Another example in biology is given
by metabolic networks. These networks are also directed and skewed. In this
case, a large number of substrates (the nodes of the graph) are involved in a
few metabolic reactions (the links) while a tiny fraction of substrates
participate in a high number of reactions \cite{metab}.

On the other hand, in the absence of conclusive experimental results, it is
difficult to know what the interaction rules of, for instance, genetic networks
actually are, although several experiments have proved that regulatory gene
networks are highly nonlinear dynamical systems \cite{nl01,nl02}. This makes it
clear that one should deal with both dynamical and structural complexity.
Recently \cite{us}, we have studied the chaotic dynamics of a continuous
gene-expression like model coupled to a complex heterogeneous network. In this
paper, we fully characterize the different dynamical regimes observed.
Specifically, we study numerically the steady, periodic and chaotic states that
appear upon variation of the system's parameters. The results obtained allow us
to draw interesting conclusions about the robustness (hereafter intended as the
ability of the system to avoid the phase space of chaotic dynamics) and
behavioral richness of complex biological networks.

The rest of the paper is organized as follows. In section \ref{model},
we describe the network's construction, introduce the model and explain
the numerical procedure. Next, the different dynamical regimes are shown
and analyzed in section \ref{dynamics}. Finally, the last section rounds
off the paper by discussing our results and giving the conclusions of
the present study.

\section{The Model}
\label{model}

The model we will discuss in what follows is built in two layers. The first one
refers to the topology of the underlying network while the second ingredient
has to do with the dynamics of the network's components. As noted before, the
topology of two relevant biological networks has been recently shown to be very
heterogeneous. This characteristic is shared by other networks in biology
\cite{book2}.  In addition, they are directed.  Henceforth, we assume that each
vertex of the underlying network corresponds to a biological entity and that
the links stand for their interactions.

We construct the underlying network in the following way. Let $C_{ij}$ be the
connectivity matrix of an undirected network built up following the
Barab\'{a}si and Albert model \cite{bar99}. This recipe allows the generation
of random scale-free networks with a degree distribution $P(k)\sim k^{-3}$ and an
average connectivity $\langle k\rangle=2m$, being $m$ the number of new links added at each
time step during the generation of the network (henceforth $m=3$ and $\langle k\rangle=6$).
The elements of the matrix $C_{ij}$ are equal to $1$ if nodes $i$ and $j$ are
connected and zero otherwise. Then, we transform $C_{ij}$ into the new matrix
$W_{ij}$ describing directional interactions \cite{notedi}. To this end, we
look over the nonzero elements of $C_{ij}$ and with probability $p$ consider
that the interaction $i\gets j$ is inhibitory, $W_{ij}=-1$, and with probability
$(1-p)$ it is excitatory, $W_{ij}=1$. Note that now the resulting matrix
$W_{ij}$ is not, in general, symmetric anymore. In this way, the parameter $p$
controls the average output (input) connectivity of each node.

The second layer of the model has to do with the individual dynamics of each
node in the underlying network. There is no model that incorporates all known
facts about a given biological process and represents efficiently and
accurately its complexity. Therefore, the development of a simplifying model is
often essential in trying to understand the phenomenon under consideration.
Here, we study a generic class of dynamical system that often appears in the
biological context and discuss the results for two plausible biochemical
processes, gene expression and reaction kinetics.

Consider that the activity of the nodes is described by the vector ${\bf
  G}(t)=\{g_1(t),g_2(t),\ldots,g_N(t)\}$, where $g_i$ ($i=1,\ldots,N$) accounts for the
activity level of each individual node $i$ in a network made up of $N$
elements. The time evolution of ${\bf G}(t)$ is described by the following set
of first-order differential equations \cite{book2,sole},

\begin{equation}
\frac{d{\bf G}(t)}{dt}=-{\bf G}(t)+{\bf F}({\bf G}(t)),
\label{eq1}
\end{equation}
where ${\bf F}({\bf G}(t))$ is some nonlinear term where the interactions
between the network's elements are taken into account. Equation\ (\ref{eq1})
includes continuous versions of Random Boolean Networks \cite{mestl,glass} as
well as continuous-time Artificial Neural Networks \cite{hopfield}, both widely
used to model periodic and chaotic dynamics in some biologically relevant
situations. Additionally, we implement a continuous Michaelis-Menten
description \cite{book2,sole,murray},

\begin{equation}
F_i({\bf G}(t))=\delta\frac{\Phi[h\sum_{j=1}^{k_i}
W_{ij}g_j(t)]}{1+\Phi[h\sum_{j=1}^{k_i}W_{ij}g_j(t)]},
\label{eq2}
\end{equation}
where $W_{ij}$ is the interaction matrix introduced before.  Additionally,
$\delta>0$ and $h>0$ are constants, $k_i$ is the connectivity of node $i$, and the
function $\Phi(z)$ is defined as follows
\begin{equation}
\Phi(z) = \left\{ \begin{array}{ll}
0  & \;\;\;\;\mbox{if $z \leq 0$} \\
z  & \;\;\;\;\mbox{if $z > 0$}
\end{array}
\right.
\end{equation}
We have set $\delta=3$ hereafter and varied $h$. One can easily realize that the
solutions ${\bf G}(t)$ for non-negative initial conditions remain bounded for
all $t>0$: As $F_i({\bf G}(t))$ is bounded above by $\delta$, $dg_i(t)/dt < 0$
whenever $g_i(t) > \delta$. Also, if $g_i(t) = 0$ then $dg_i(t)/dt = F_i({\bf
  G}(t))\geq 0$, so that the activities cannot become negative.

The dynamics of the system defined as before is determined by only two
parameters, $h$ and $p$. One controls the degree of nonlinearity and the other
the topological properties of the network, respectively. We have performed
extensive numerical simulations of the set of equations\ (\ref{eq1}-\ref{eq2}).
Starting from small values of $h$, the time evolution of the local dynamics
$g_i$ is obtained by means of a 4$th$-order Runge-Kutta integration scheme
\cite{recipe}. The set of simulations carried out screens the parameter space
$(h,p)$, where $h$ goes from $1$ to $10$ and $p$ from $0$ to $1$. For each pair
$(h,p)$, different realizations corresponding to many initial conditions and
network realizations were performed.

This dynamics turns out to be very rich and, depending on the values of $p$ and
$h$, three different asymptotic dynamical regimes are observed, characterized
by stationary, periodic and chaotic attractors.  All three states may even
coexist in a given network realization, each in different islands or clusters.
Islands are subnetworks that are interconnected through nodes which have
evolved to null activity, and so (asymptotically) their dynamics are
effectively disconnected.

While stationary and periodic states point to regions of the parameter space
where real biological networks might operate, the existence of chaotic dynamics
would be, in general, inconsistent with the reproducibility of experimental
observations in living organisms. Hence, we have characterized all possible
responses of the system under variations of both $h$ and $p$ and monitored the
evolution of $g_i(t)$, the probabilities of ending up in either chaotic or
periodic dynamics as well as the distribution of clusters or islands of nodes
displaying such behaviors. Moreover, due to recent interest in what is known as
network motifs, we have also analyzed the topological features of the clusters
exhibiting non-stationary behavior.

The computations presented in the rest of the paper were developed
following this sequence:

\begin{itemize}

\item[({\it i})] The initial values of $g_{i}$ are taken from a
uniform distribution in the interval $(0,1)$.

\item[({\it ii})] First integration of the equations is performed
using a $4^{th}$ order Runge-Kutta scheme \cite{recipe}. The total
integration time is large compared with the transient.

\item[({\it iii})] Check the dynamical state of the network. If all
the nodes are in a steady state we try another initial configuration;
if there are dynamical nodes go to the next stage.

\item[({\it iv})] Check the connectivity between the dynamical nodes
in order to obtain the dynamical subnetworks (islands).

\item[({\it v})] Second integration for calculating the largest
Lyapunov exponent $\lambda$ \cite{book4}. If $\lambda>5\cdot 10^{-3}$
the dynamics is considered chaotic. If $\lambda<5\cdot 10^{-3}$ we
look at the frequency of the periodic motion.

\item[({\it vi})] Repeat stages ({\it i})-({\it v}) for different
initial conditions and realizations of the network.

\end{itemize}

We have generated networks of sizes ranging from $N=100$ to $N=300$
nodes. At each value of $p$ and $h$, we have performed at least $1000$
iterations of the above procedure. The time step in the integration
scheme was fixed to $10^{-4}$. We incorporate later on a further
criterion in order to obtain the values of the frequencies of the
periodic states.

\section{Dynamical regimes}
\label{dynamics}

\begin{figure}
\begin{center}
\begin{tabular}{cc}
\resizebox{8cm}{!}{
\includegraphics[angle=-90]{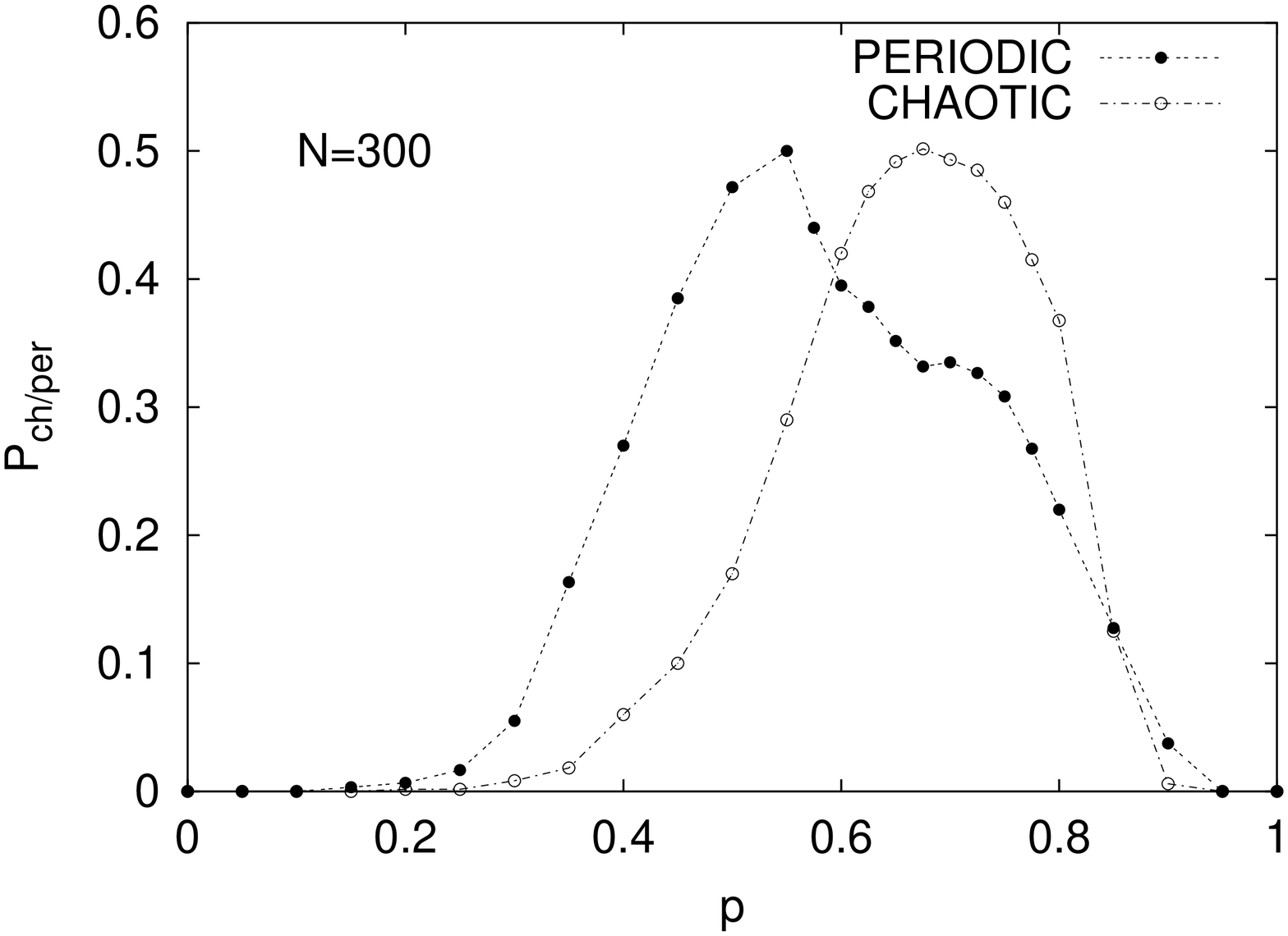}
}
\end{tabular}
\end{center}
\caption{Probabilities of having chaotic, $P_{ch}$ or periodic
  $P_{per}$ behavior as a function of the topological parameter $p$. The value
  of $h$ has been set to $4$ and the system is made up of $N=300$ nodes. See
  the text for further details on the definitions.}
\label{figure1}
\end{figure}

The individual dynamics of the nodes is not uniform across the entire network
due to the heterogeneity in the initial conditions and that of the underlying
networks. While some nodes reach a stationary state, others follow periodic
orbits and even chaotic behavior. The following argument explains how this can
easily happen. If a node $i$ is such that $W_{ij} =-1$ for all $j$, then its
activity will tend to zero. The same will happen for those nodes $l$ such that
the positive $W_{lj}$ occur for $j$'s of the previous kind, etc{\ldots} Now, two
subnetworks connected between them through nodes whose activity dies out will
become effectively disconnected, and so their dynamics are asymptotically
independent. The network is then dynamically fragmented into islands.

A simple way in which the overall dynamics of the network can be described is
through the computation of the largest Lyapunov exponent $\lambda$. Once we have
obtained $\lambda$, we define the probability that a given dynamical regime is
observed. As they are complementary, we define only two probabilities. Namely,
the probability that the network displays chaotic behavior, $P_{ch}$, is the
fraction of the total number of realizations in which at least one node ends up
in a chaotic state yielding a positive value of $\lambda$. On the other hand, if $\lambda\leq
0$, the system does not end up in a chaotic regime and only stationary and/or
periodic islands are observed. Consequently, the probability that no chaotic
behavior is attained, but periodic orbits are observed, $P_{per}$, is given by
the portion of the total number of realizations in which $\lambda \leq 0$ and there is
at least one periodic orbit.

Figure\ \ref{figure1} shows the two probabilities as a function of the
topological parameter $p$ for a fixed value of $h=4$ and $N=300$. Two different
threshold values for $p$ can be observed, mainly determined by $P_{per}$, $p_1\approx
0.2$ and $p_2\approx 0.95$. The first region $p < p_1$ corresponds to the case in
which most of the interactions are excitatory and the individual dynamics are
described by frozen steady states (either $g_i=0$ or $g_i\geq 0$). On the other
hand, when the interactions become predominantly inhibitory ($p > p_2$), the
activity of the nodes dies out due to the damping term in Eq.\ (\ref{eq1}). In
the intermediate region, $p_1 < p < p_2$, all types of behaviors (stationary,
periodic and chaotic) are achieved.

\subsection{Stationary states}

The simplest case of stationary state is the solution in which all
the nodes remain inactive, {\em i.e.} $g_i = 0$ for all $i$. Note
that this is always a solution of the equations of motion,
irrespective of the parameter values. As a matter of fact, for
$h=0$, or $h\neq0$ but $p=1$, the state of inactivity (or rest
state) is the unique asymptotic solution for any non-negative
initial conditions. However, for $h\neq0$ and $p\neq1$ other
asymptotic solutions with islands of positive activity generically
coexist.

\begin{figure}
\begin{center}
\begin{tabular}{cc}
\resizebox{7.cm}{!}{
\includegraphics[angle=-90]{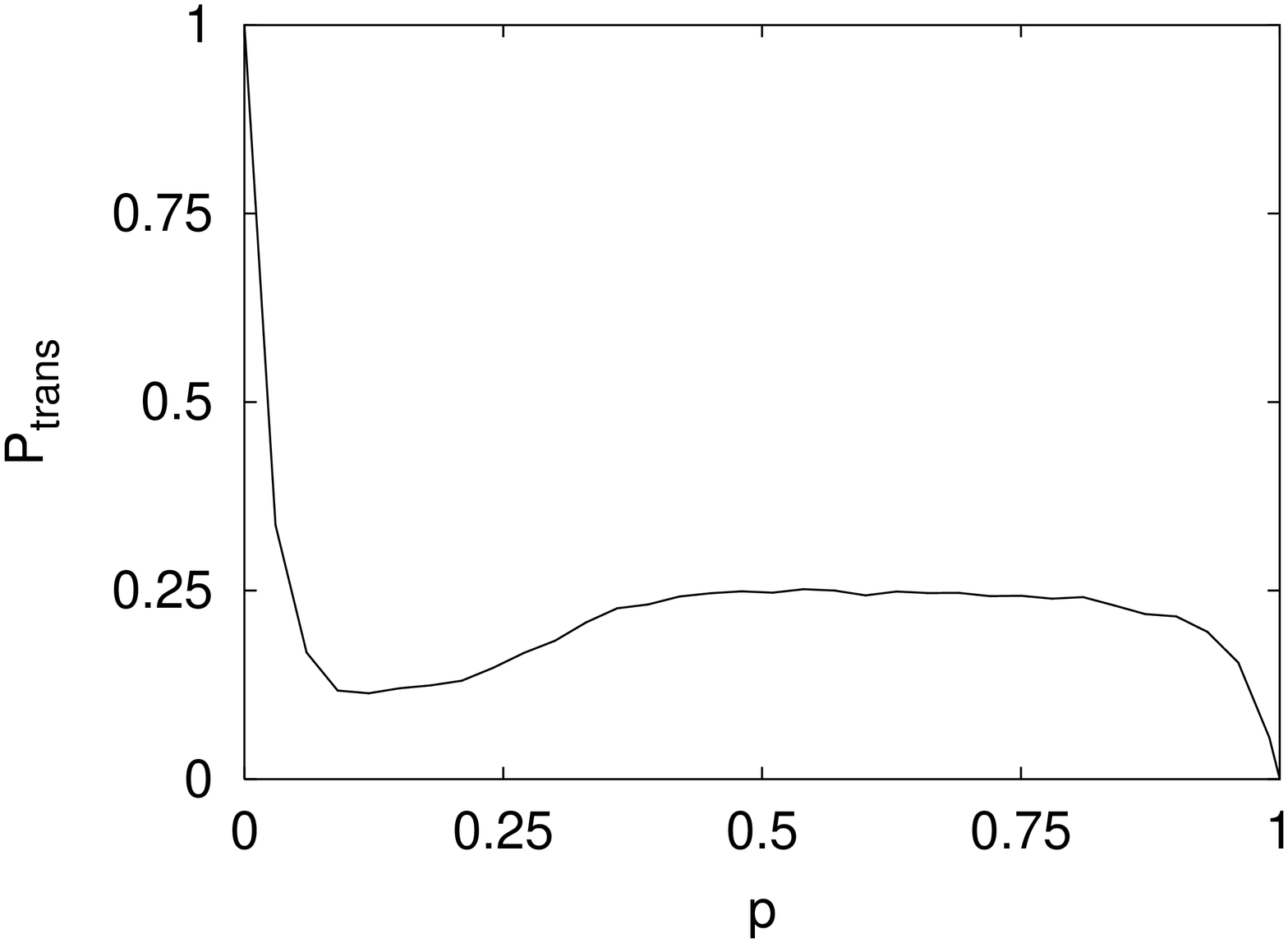}
}
\end{tabular}
\end{center}
\caption{Probability $P_{trans}$ that the uniform state of node 
inactivity becomes unstable for some value of $h$, as a function 
of the parameter $p$. $10^4$ different realizations have been used 
for each value of $p$.} \label{figPtrans}
\end{figure}

Depending on the specific network realization ({\em i.e.} the matrix $W_{ij}$),
the rest state can become unstable when the value of $h$ is increased from
zero. This will occur for the value $h=\tilde{h}$ at which the largest
eigenvalue (among those associated to eigenvectors such that all their
components have the same sign \cite{note0}) of the matrix $-\delta_{ij} + \delta h
W_{ij}$ becomes positive. Then $\tilde{h}$ is determined as $1/(\delta \lambda_{max})$,
where $\lambda_{max}$ is the largest eigenvalue of $W_{ij}$, provided $\lambda_{max}>0$ (no
instability of the rest state will occur if $\lambda_{max}\leq0$). In Fig.\ 
\ref{figPtrans} we show the probability $P_{trans}$ that the rest state becomes
unstable for some value of $h$, as a function of the parameter $p$. This
probability has been estimated from the computation of $\lambda_{max}$ for $10^4$
different realizations of $W_{ij}$ for each value of $p$. Though for most
values of $p$ the rest state remains stable at all values of $h$, in $75\%$ (or
more) of the realizations, it coexists in phase space with other attractors, so
that only a basin of initial conditions evolves to this state.

\begin{figure}
\begin{center}
\resizebox{14.cm}{!}{
\includegraphics[angle=-0]{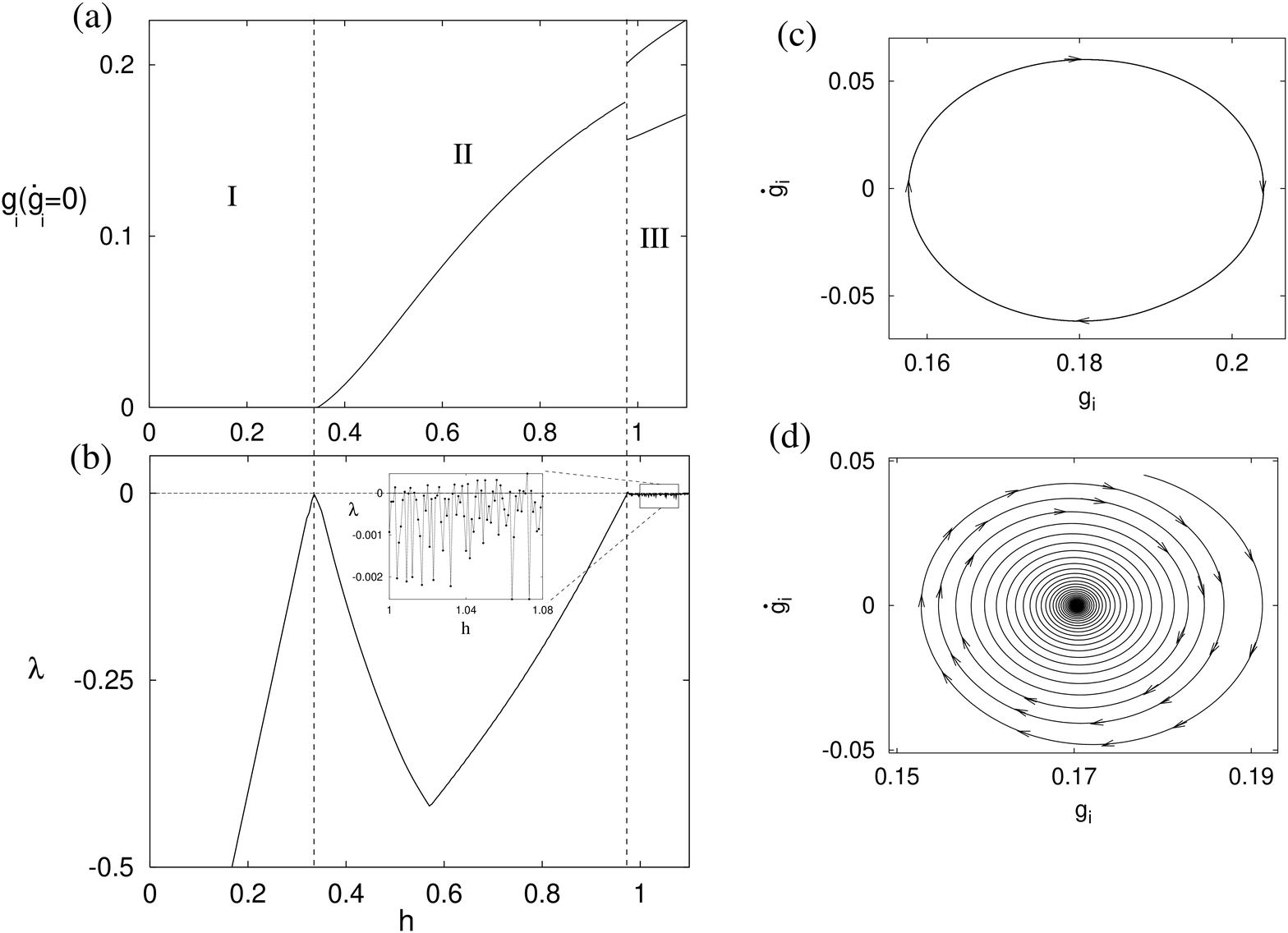}
}
\end{center}
\caption{{\bf (a)} Dependence of the quantity $g_{i}[\dot{g}_{i}=0]$ (activity
  level when its first derivative is zero) of a single node with the parameter
  $h$. This node belongs to a cluster which undergoes two bifurcations when
  increasing the value of $h$ starting from the rest state configuration of the
  whole network at $h=0$.  In the first bifurcation ($h\simeq0.345$) a cluster of
  nodes in a stationary state with non-zero activity level merges. When
  $h\simeq0.976$ the nodes of this cluster end in a periodic attractor.  The
  evolution of the largest Lyapunov exponent of the network as $h$ is increased
  is plotted in {\bf (b)} showing the two bifurcations.  {\bf (c)} Periodic
  trajectory in the portion of the phase space corresponding to the node of
  figure {\bf (a)}, the value of $h$ is $0.98$. {\bf (d)} Decay of the activity
  level of the same node to the stable fixed point for $h=0.97$ (just before
  the second bifurcation) when the initial condition of the network is the
  periodic solution shown in {\bf (c)}.  }
\label{figstationary}
\end{figure}

The rest state destabilizes typically through a transcritical
bifurcation \cite{Seydel}, where an unstable branch of stationary
solutions exchanges stability with the rest branch (see
Fig. \ref{figstationary}a). The computed largest Lyapunov exponent
shows a variation with $h$ as in Fig. \ref{figstationary}b near
$h\simeq0.345$: it approaches zero (from negative values) at the
bifurcation parameter value, and then decreases indicating that now
the attractor belongs to the new stable stationary branch, in which
the nodes of a cluster display non-zero activity. As shown in
Fig. \ref{figstationary}a, the activity of these nodes typically
increases with $h$. Eventually, this state becomes unstable for larger
values of $h$, typically through a Hopf bifurcation (either inverse or
direct) to a periodic state in which the activities oscillate (Fig.\
\ref{figstationary}c) regularly in time.

\subsection{Periodic states}

The observation that both $P_{per}$ and $P_{ch}$ in figure \ref{figure1} are
zero outside the interval $[p_1,\; p_2]$ of values of the parameter $p$ clearly
indicates that non-stationary activity is the result of the interplay between
excitatory and inhibitory interactions in the network.

\begin{figure}
\begin{center}
\begin{tabular}{cc}
\resizebox{8cm}{!}{
\includegraphics[angle=-90]{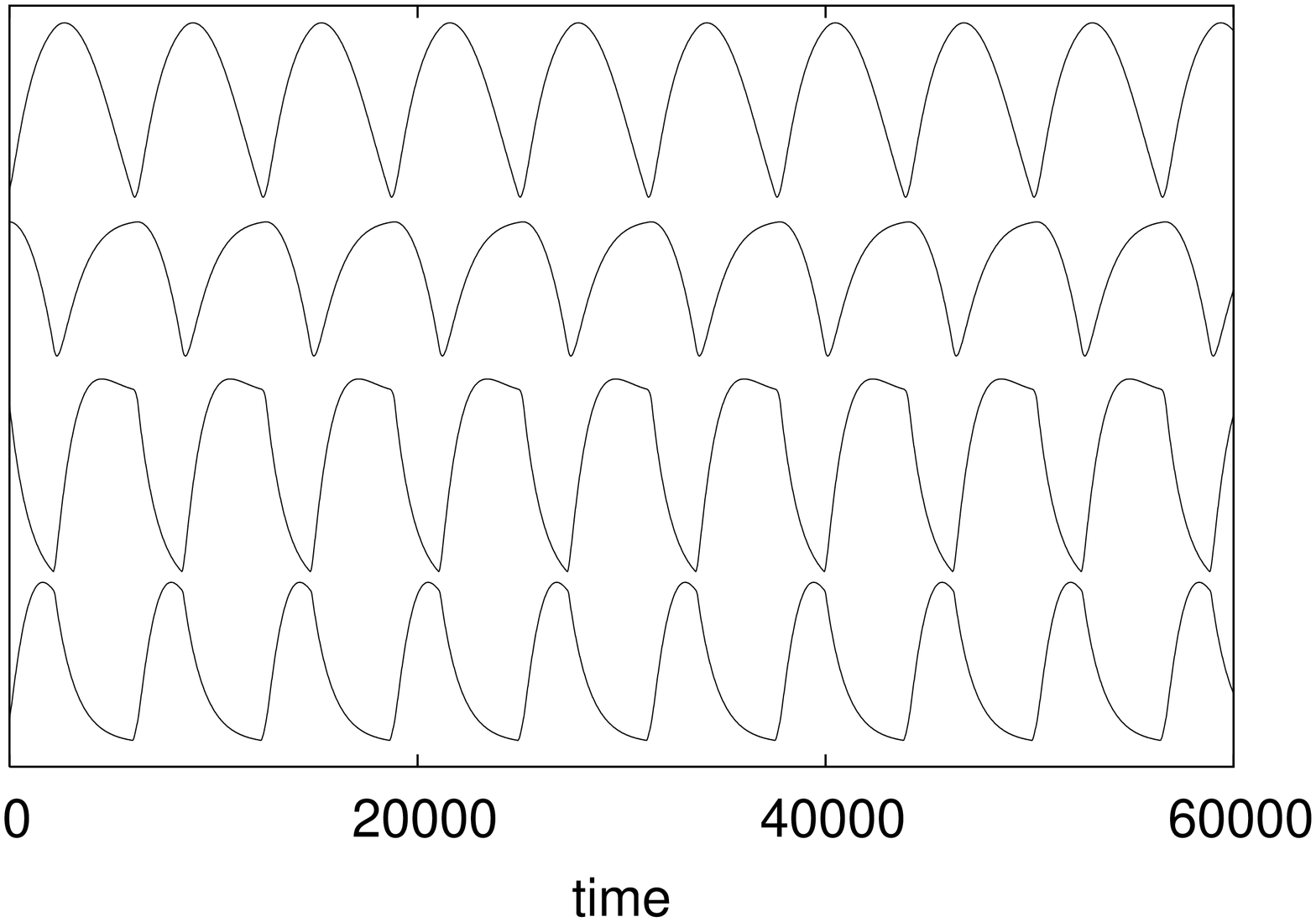}
}
\end{tabular}
\end{center}
\caption{Time series for four different nodes exhibiting periodic
  dynamics. Time is in dimensionless units and starts just after the
  transient period. The value of $p$ is $0.7$ and $N=100$.}
\label{figure6}
\end{figure}

\begin{figure}[t]
\begin{center}
\begin{tabular}{cc}
\resizebox{8cm}{!}{
\includegraphics[angle=-90]{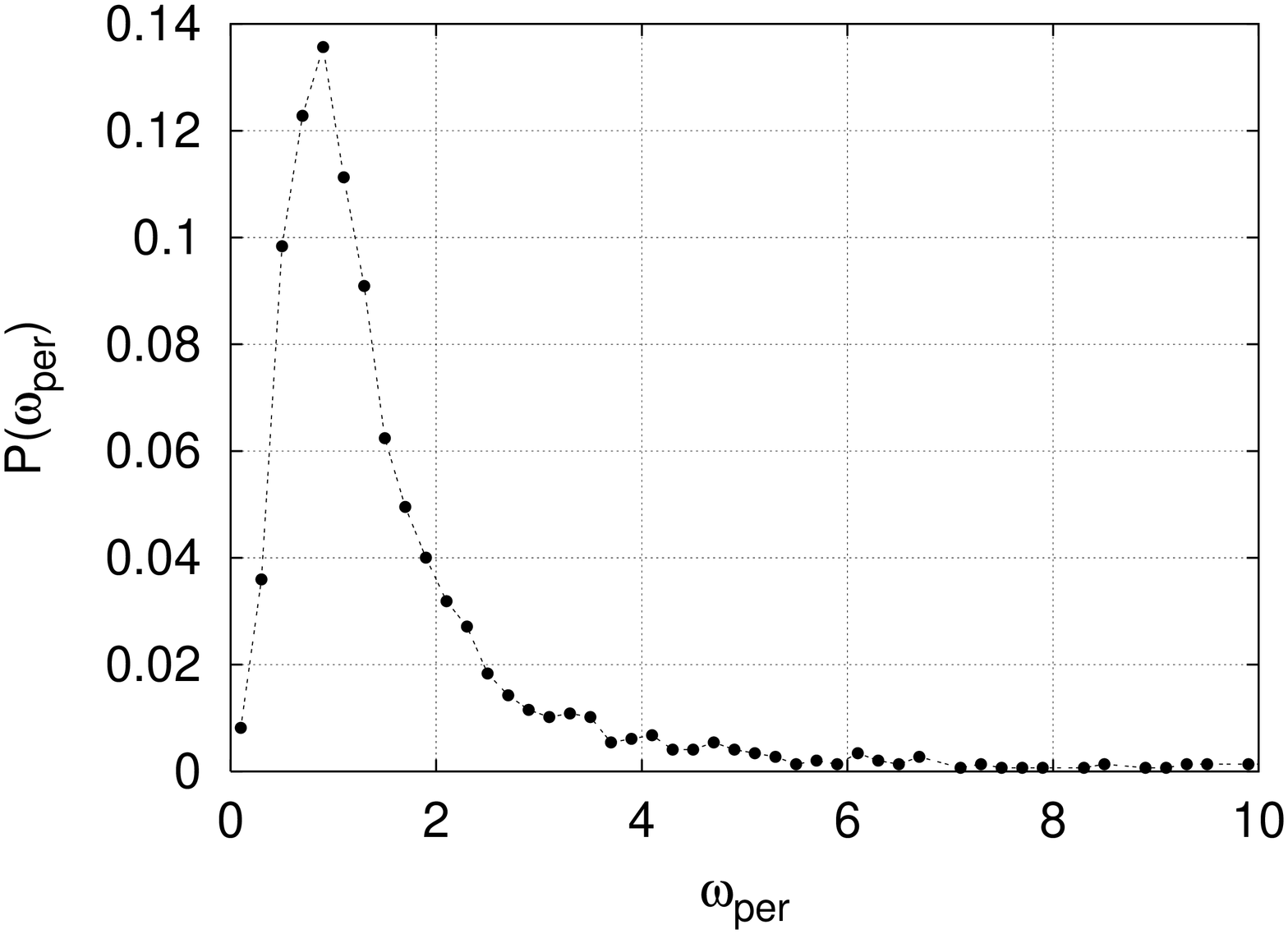}
}
\end{tabular}
\end{center}
\caption{Probability that a node with periodic dynamics converges to
  an orbit of angular frequency $\omega_{per}$ (in arbitrary
  units). The results are averaged over different network realizations
  and at least 100 different initial conditions for a network of
  $N=100$ nodes and $p=0.7$. $h$ has been fixed to $4$.}
\label{figure7}
\end{figure}

In Fig.\ \ref{figure6}, we have plotted the time profiles of four different
nodes in a periodic regime within the same island. One observes out of phase
oscillations which reflect the existence of inhibitory interactions: the growth
of the activity in the node $j$ inhibiting node $i$ ($W_{ij}=-1$) leads
eventually to a null value of $F_i({\bf G}(t))$, thus to an exponential (free)
decay of the activity of node $i$, until it is triggered again (due to the
decay of the activity of inhibitory interactions and/or the increase of the
excitatory nodes' activity). As the free decay of activity has an associated
time scale of order unity, one should expect values of this order for the
period of oscillations.

This expectation is confirmed by computing the frequency distribution of nodes
whose dynamics converge to a periodic cycle for different realizations. The
numerical procedure is as follows: First, we identify the realizations in which
the largest Lyapunov exponent is zero. Then, we focus on the nodes for which
$dg/dt\neq 0$.  Once identified, a vector ${\bf T}_n^i=\{t_1^i,t_2^i,\ldots,t_n^i\}$ is
constructed and stored for every periodic dynamics $g_i$. The $t_j^i$'s stand
for the times fulfilling the conditions $g_i(t_1^i)=g_i(t_2^i)=\ldots=g_i(t_n^i)$
and $dg_i(t_1^i)/dt=dg_i(t_2^i)/dt)=\ldots=dg_i(t_n^i)/dt$ \cite{note1}. In this
way, after verifying that $t_j^i-t_{j-1}^i$ is constant, the period of the
corresponding $i$-orbit is given by this constant.

In Fig.\ \ref{figure7}, we show the probability that a periodic cycle has an
angular frequency $\omega_{per}$. As shown in the figure, it is very likely that the
frequency of the activity of a periodic island lies around $\omega_{per}=1$. It is
also of interest that $P(\omega_{per})$ is not symmetric, but biased towards larger
frequency values. It is difficult to figure out an explanation to this
behavior. It may probably have to do with the spatial distribution of the nodes
and the specific value of $p$ which controls the average number of input and
output connections a node has.

\subsection{Chaotic states}

\begin{figure}
\begin{center}
\begin{tabular}{cc}
\resizebox{8cm}{!}{
\includegraphics[angle=-90]{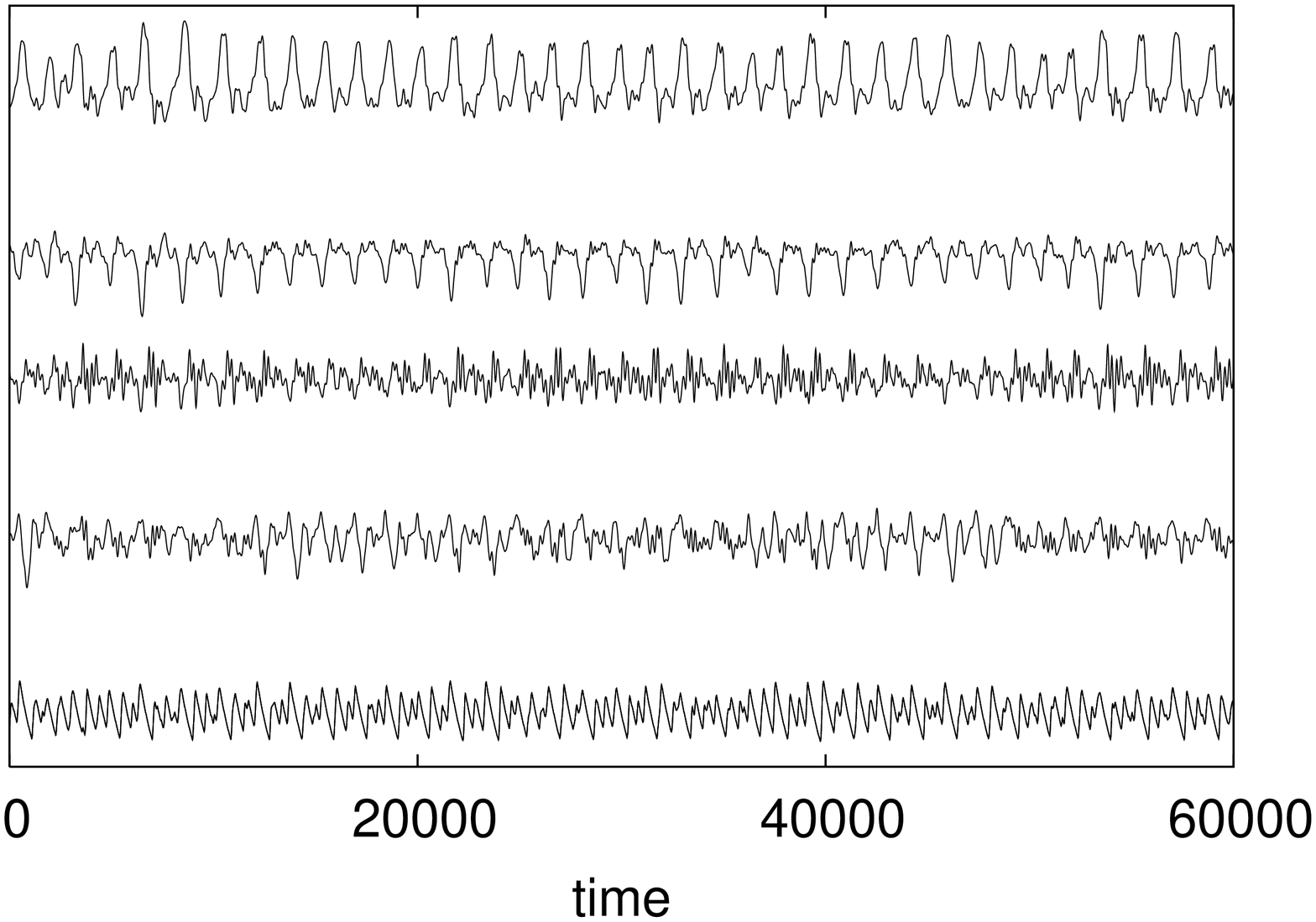}
}
\end{tabular}
\end{center}
\caption{Time series for five different nodes exhibiting chaotic
  dynamics. Time is in dimensionless units and starts just after the
  transient period. The value of $p$ is $0.7$ and $N=100$.}
\label{figure2}
\end{figure}

Although in general not desirable from a biological point of view, systems
displaying chaotic behavior are always of interest \cite{book4}. Moreover, the
existence of chaotic dynamics does not only depend on parameters associated to
the dynamics employed (as in most of the studies performed so far regarding
chaotic dynamics), but more important, it is the result of a complex interplay
between the dynamical and structural (topological) complexity. We next
summarize the results obtained for the chaotic dynamics of the system.

The two threshold values in the phase diagram for the chaotic regime depicted
in Fig.\ \ref{figure1} depends on $h$. Clearly, as the degree of nonlinearity
increases, chaotic behavior appears more frequently, which translates in a
larger maximum for $P_{ch}$. On the other hand, although we have used a small
system size, the values of $p_1$ and $p_2$ seem to be robust when $N$ grows.
This means that the results obtained are meaningful for larger systems since
the onset and the end of the chaotic phase are $N$ independent. In Fig.\ 
\ref{figure2} we have represented the time profile of five different nodes in
the chaotic regime.  Time units refer to integration steps and the origin of
the time scale begins just after the transient period. The system's parameters
are as indicated. Note that although all these nodes are in the same chaotic
island, the patterns of activity are quite different and the amplitudes of the
chaotic signal (and the shape of the curves) are distinguishable.

\begin{figure}
\begin{center}
\begin{tabular}{cc}
\resizebox{12cm}{!}{
\includegraphics[angle=-0]{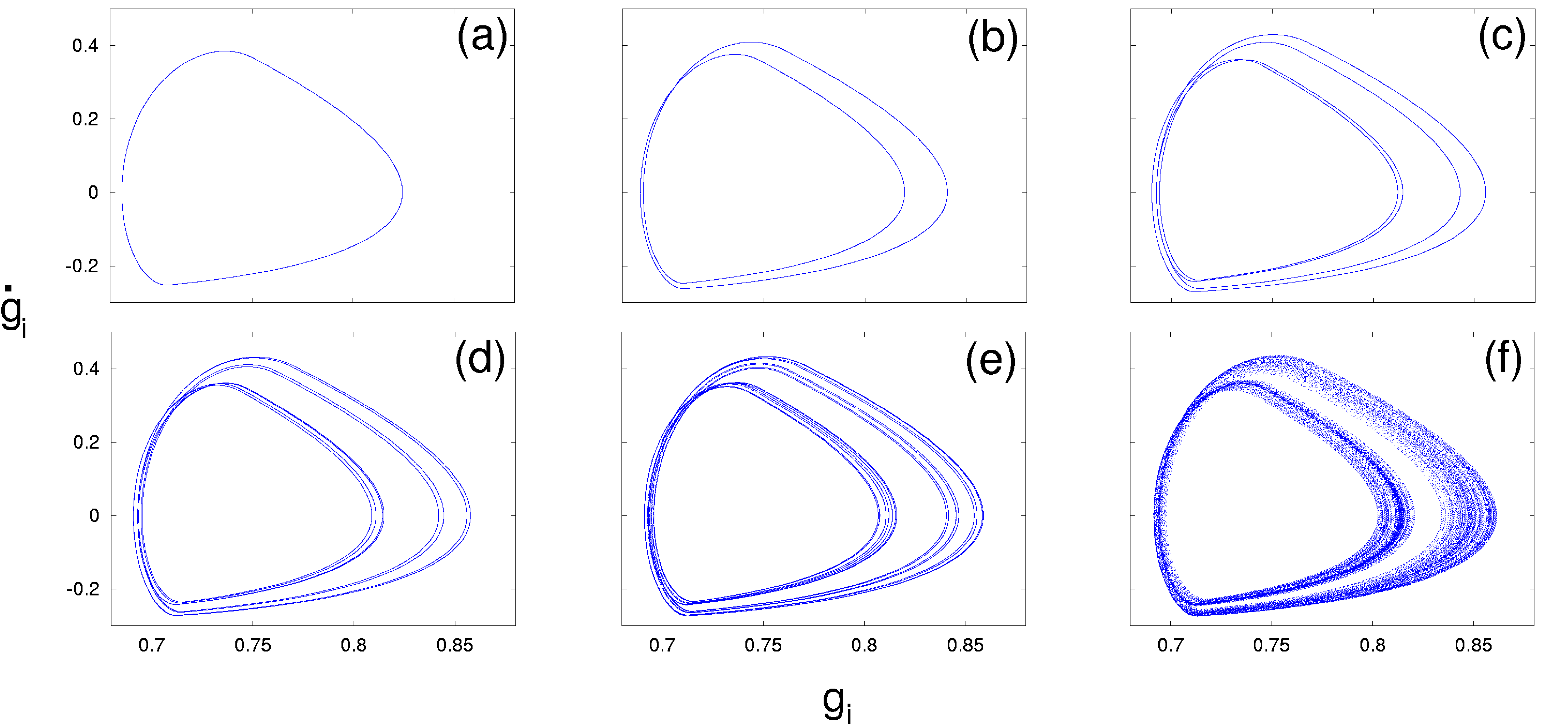}
}
\end{tabular}
\end{center}
\caption{Phase space of a node ending up in a chaotic state as the
  value of $h$ is increased. Successive period doublings starting from a
  periodic cycle can be appreciated. The values of $h$ (from {\bf (a)} to {\bf
    (f)}) are: $5.30$, $5.50$, $5.63$, $5.65$, $5.66$, $5.68$, respectively.
  Other network parameters are as in Fig.\ \ref{figure2}.}
\label{figure5}
\end{figure}

It is also of interest to know how the chaotic regime is attained.  The origin
of different dynamical patterns is related to the values of $p$ and $h$. First,
we study the transition to chaos from periodic states. We have traced the route
to chaos \cite{us} by picking up a node at random among the chaotic ones. By
increasing the value of $h$ at intervals of $\Delta h=0.02$, we recorded the local
maxima of $g_i$ in the corresponding time series. The results reveal that the
chaotic regime is reached through the period-doubling cascades mechanism
\cite{us}.

As an evidence of the period-doubling mechanism, Fig.\ \ref{figure5} shows the
phase space diagrams of the node's activity as $h$ is increased. For small
values of $h$, the node is in a periodic cycle, which doubles its period
successively until it reaches the chaotic phase. This corroborates that when
$h$ and $p$ allows for a large value of $P_{ch}$, the behavior of the system is
dominated by dynamical states (either periodic or chaotic). Moreover, the fact
that $\sum P_{ch}+P_{per}$ is large indicates that nodes which are not in a
chaotic state may be in the route to it. In other words, in this regime of
parameters, when a given realization has no chaotic islands, it is very likely
that it has periodic clusters.

Up to now, we have described the activity patterns in terms of their dynamical
properties. However, one of the most interesting aspects of many biological
networks is that their topology is highly heterogeneous. There are a few nodes
which interact with many others. This should have some bearings in the results
obtained. In the next section, we round off our numerical analysis of the model
by correlating the dynamical properties unraveled with the network's
topological features.

\subsection{Structural properties of dynamical regimes}

\begin{figure}
\begin{center}
\begin{tabular}{cc}
\resizebox{8cm}{!}{
\includegraphics[angle=-90]{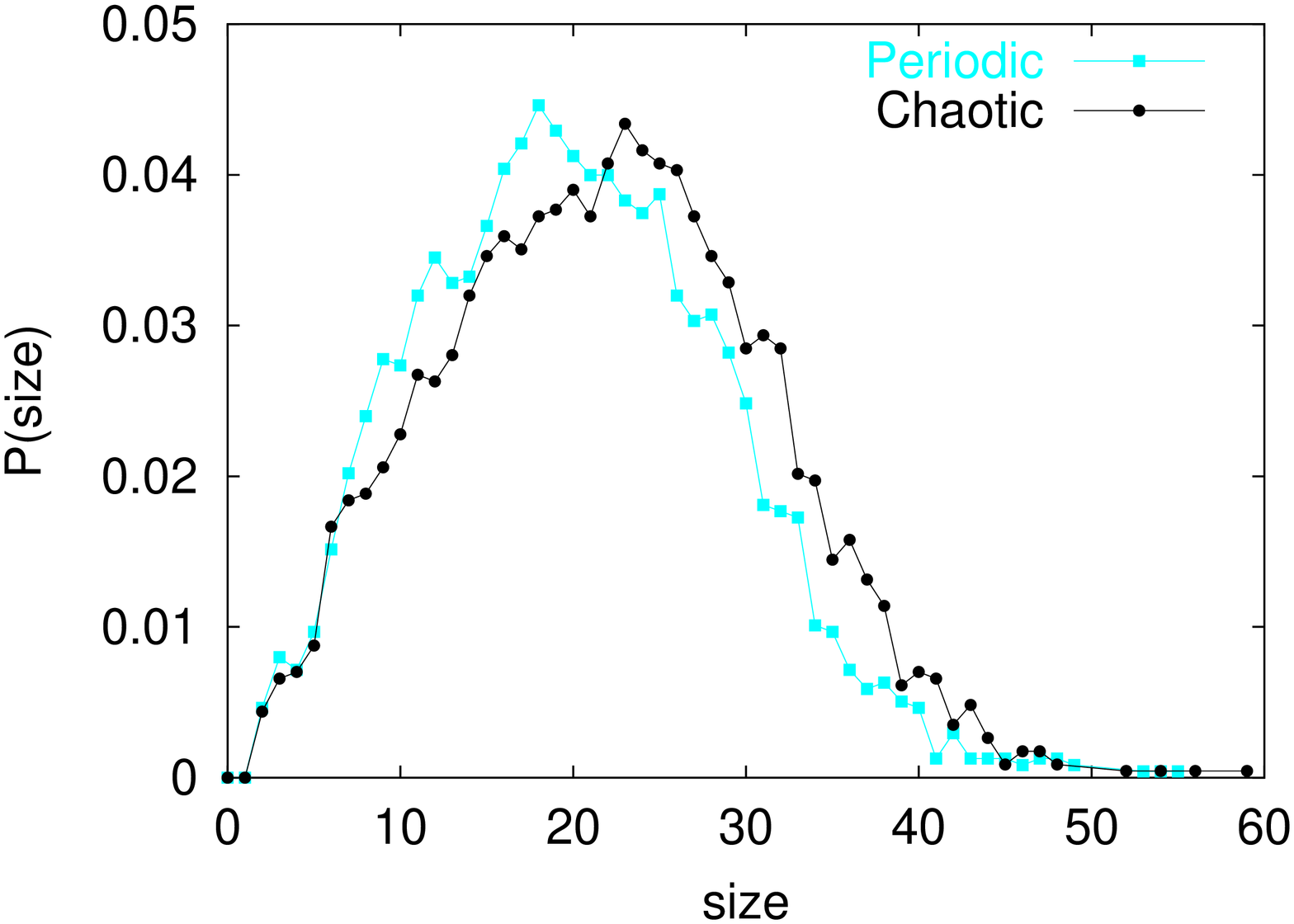}
}
\end{tabular}
\end{center}
\caption{Probability that a connected cluster of nodes displaying
  either chaotic or periodic behavior has a given size (in number of
  nodes forming the cluster). Although the curves correspond to two
  radically different behaviors, they are very similar. Network
  parameters are as in Fig.\ \ref{figure7}.}
\label{figure8}
\end{figure}

In order to characterize the structural properties of distinct dynamical
regimes, we focus on some magnitudes. The first and simplest structural
characterization is given by the distribution of nodes exhibiting either
periodic or chaotic dynamics, {\em i.e.}  the histograms of periodic (and
chaotic) cluster sizes.

\begin{figure}[t]
\begin{center}
\begin{tabular}{cc}
\resizebox{10cm}{!}{
\includegraphics[angle=-0]{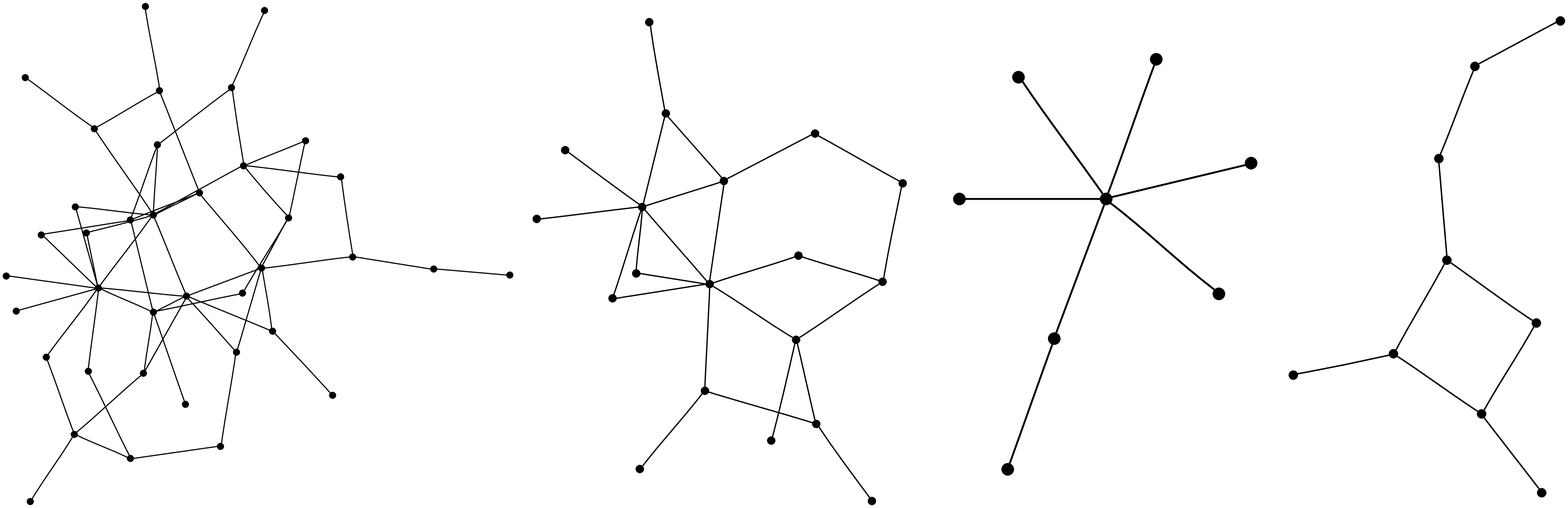}
}
\end{tabular}
\end{center}
\caption{Examples of chaotic clusters (or islands) whose size
distribution is reported in the previous figure. As one can see, no
typical structure arises regardless of the existence of a well-defined
average cluster size.}
\label{figure9}
\end{figure}

The results are shown in Fig.\ \ref{figure8}, where the size distribution of
the two dynamical behaviors are drawn for $N=100$ elements, $p=0.7$, and $h=4$.
Apart from slight fluctuations in the maxima of both curves, it is apparent the
existence of a mean average cluster size for chaotic and periodic islands,
though the dispersion around this mean value is relatively large. The fact that
both curves almost collapse into a single one indicates that the clusters of
periodic nodes are the same that later on, by increasing $h$ at fixed $p$,
evolve to a chaotic state. Moreover, since the largest clusters are made up of
roughly half of the network's constituents, it is highly improbable that the
entire system displays the same behavior. In other words, the fragmentation of
the network into islands of independent dynamics appears as one of the most
characteristic features of this model. In Fig.\ \ref{figure9}, we show four
different clusters corresponding to nodes with chaotic dynamics. As one can
see, no typical structure appears, even for clusters of comparable sizes,
except that all of them have a relatively small value of the cluster average
connectivity.

In addition, we would like to add a few sentences about network motifs, a
subject that has become of utmost interest in biological and other networks due
to its implications in modularity and community structures \cite{modularity}.
Motifs are small graph components or loops that appear more frequently than in
a random network with identical degree distribution \cite{book2,milo2}. We have
found no correlation between the clusters of periodic or chaotic behavior with
the structural properties of the underlying network. In particular, chaotic or
periodic clusters do not follow a universal topological pattern, contrary to
what one may expect from other studies on network evolution and motifs (see,
for instance, Maslov et al. in ref.\ \cite{book2}, and \cite{evol}). We believe
that this is due to the fact that the underlying network, although
heterogeneous and with directed interactions, is a random scale-free network,
for which the probability that a motif exists is much lower than in real
biological networks. 

\begin{figure}
\begin{center}
\begin{tabular}{cc}
\resizebox{10cm}{!}{
\includegraphics[angle=-90]{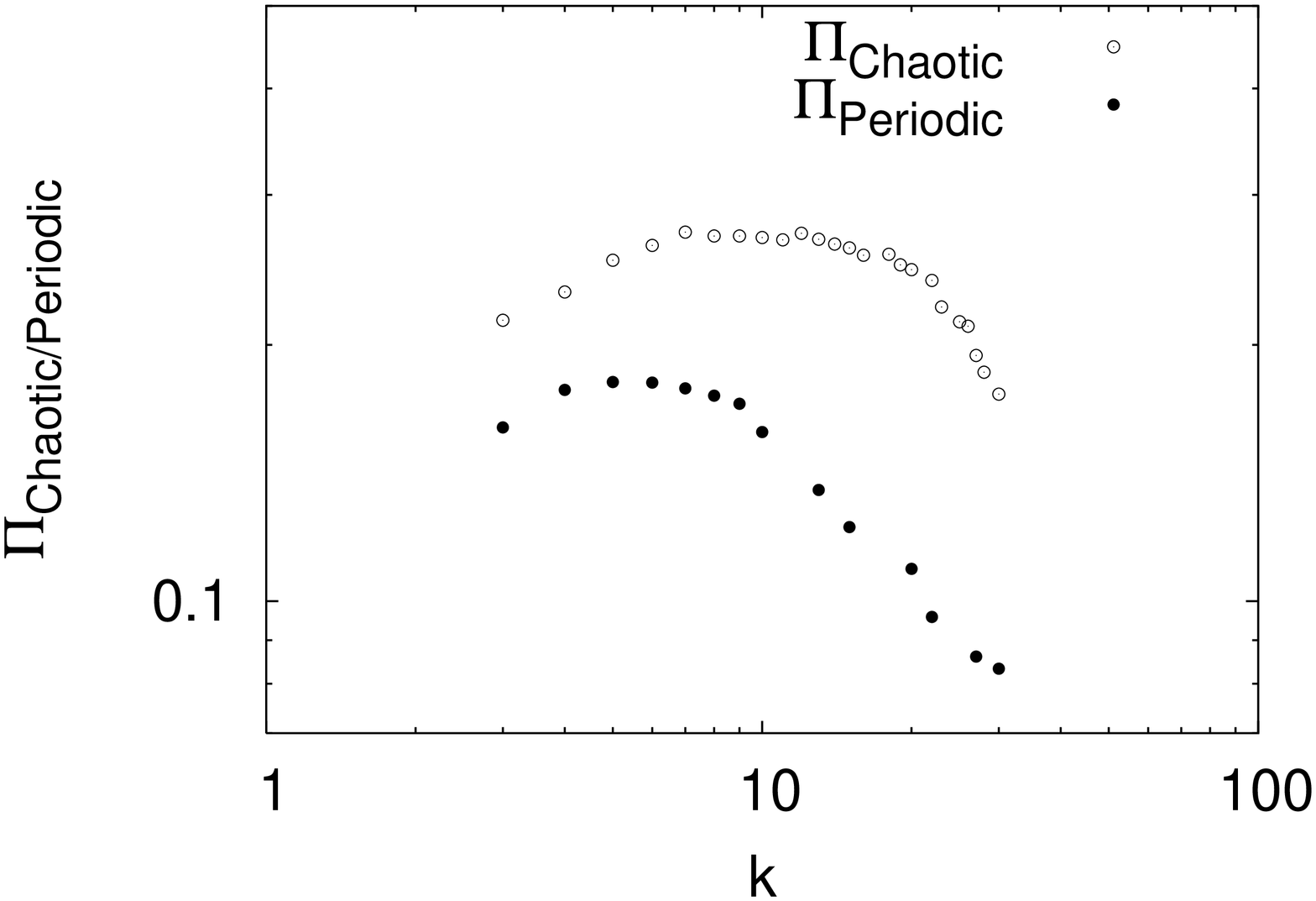}
}
\end{tabular}
\end{center}
\caption{Probability of having periodic or chaotic behavior in a node of
  connectivity $k$. The figure indicates that dynamical regimes are
  well differentiated by connectivity classes. Network parameters are as of
  Fig.\ \ref{figure2}.}
\label{figure10}
\end{figure}

The heterogeneity of the underlying network allows us to further scrutinize the
correlations between structural and dynamical properties. In particular, it is
also of interest to elucidate what nodes take part in each regime according to
their connectivities. We have anticipated in \cite{us} that highly connected
nodes are less likely in a chaotic regime. However, due to the small size of
the networks, the results there shown are biased by the huge fluctuations in
the tail of the connectivity distribution. It is then advisable to work instead
with the cumulative distribution in order to rule out as much as possible the
noise in the tail of $P(k)$. We have monitored the probability that a node with
connectivity $k$ displays either chaotic or periodic behavior. These
probabilities, $\Pi_{chaotic}$ and $\Pi_{periodic}$ respectively, are defined as
the ratio between the number of nodes with degree $k'$ greater than or equal to
$k$ that end up in a chaotic (periodic) regime and the total number of nodes
with connectivity $k'\geq k$ averaged over many realizations. Figure\ 
\ref{figure10} shows the results obtained. 

Two interesting issues are worth mentioning. On one hand, the fact that the
chaotic phase is reached through the doubling-period cascade mechanism leads us
to expect that $\Pi_{chaotic}$ and $\Pi_{periodic}$ behave in a similar fashion,
which is confirmed in the figure.  In other words, nodes that are not chaotic
are in cycles that eventually will double their periods (as $h$ is increased)
until chaotic states are attained.  This means that if each cycle is identified
with a variety of intracellular tasks, the transition from one cycle to another
and to a chaotic phase is continuous and does not occur suddenly. On the other
hand, the results in Fig.\ \ref{figure10} points to the differences in the
nodes' dynamics according to their connectivities. In particular, apart from an
apparent maximum attained around the mean average connectivity, the results
show that the more connected a node is the less likely it is in a chaotic
regime or in the route to it (this is more clearly appreciated for
$\Pi_{periodic}$).

\section{Discussions and Conclusions}

A large number of systems have been studied in the last several years from the
network perspective \cite{book1,book2,book3}. This approach has allowed the
understanding of the effects of complex topologies in many well-studied
problems. By comparing the results obtained with other topologies and those for
real graphs in processes such as the spreading of epidemic diseases
\cite{pv01a,moreno02} or rumors \cite{maziar} and the tolerance of complex
networks to random failures and attacks \cite{newman00}, we have realized that
topology plays a fundamental role.

As for biological systems, a very rich behavioral repertoire is well documented
\cite{murray}. Cycles in biological systems range from circadian clocks to the
oscillations observed in the concentration of certain chemicals, for instance,
in biochemical reactions such as glycolysis. On the other hand, the molecular
basis for chaotic behavior have also been discussed, though chaotic behavior is
probably not a relevant issue in biology as intrinsic noise makes it difficult
to isolated truly chaotic regimes with current experimental techniques. The
dynamics studied in this work, however, could plausible describe at least to
biological scenarios, namely, gene expression and reaction kinetics in
metabolic networks.

The description of gene dynamics has to rely necessarily on models that are
only an approximation to that of real genes. However, there are several
successful approaches to the individual dynamics of gene expression. Since its
introduction by Kauffman several decades ago, one of the most studied class of
models are the Random Boolean Networks (RBNs) \cite{kauffman}, which have been
shown to lead to some predictable properties and guided our understanding
towards more complex descriptions.  Our model belongs to a generalized class of
RBNs. It takes into account the fact that genome regulation involves continuous
concentrations of RNA and proteins. The latter class has also been extensively
studied in the last years (see, for instance, Sol\'e and Pastor-Satorras in ref.\ 
\cite{book2} and references therein). In this context, each vertex would
represent a regulatory gene and the links would describe their interactions.
In other words, two nodes at the ends of a link are considered to be
transcriptional units which include a regulatory gene. One of these end-nodes
can be thought of as being the source of an interaction (the output of a
transcriptional unit). The second node represents the target binding site and
at the same time the input of a second transcriptional unit.

In genetic network models, it is known that at least for RBNs, the results are
affected by the type of Boolean functions used, the number of network's
constituents and the average connectivity (mainly input connectivity) of the
genes \cite{kauffman}. Recently, it has also been demonstrated that the degree
distribution matters \cite{oosawa}. In our continuous model, results obtained
for homogeneous random networks \cite{us} indicate that the values of $p$ for
the onset of chaos change as well. On the other hand, as pointed before, gene
networks are constrained by their required functional robustness. Therefore, as
chaos represents a long-term behavior that exhibits sensitive dependence on
initial conditions, real biological gene networks must not systematically
operate in the parameter region where the existence of chaotic attractors is
likely. By studying simplified models as the one implemented here $-$ the
intrinsic complexity of the problem does not allow for a complete and detailed
description of real gene dynamics $-$, one can infer the region of the
parameter space (i.e. $(p,h)$) that better describes gene networks. Based on
this hypothesis, by exploring magnitudes as the one represented in Fig.\ 
\ref{figure1}, we would either guess dynamical interaction rules or provide
hints for the experimental validation of the structural topology of real
networks. The latter seems more feasible due to latest developments in
microarray technologies, biocomputational tools, and data collection software.

On the theoretical side, the continuous model employed here shares some
features with respect to those seen in RBNs. For instance, although $p_1$
depends on the specific value of $h$ entering Eq.\ (\ref{eq2}), it lies in the
interval $(0.2,0.6)$ for $1 \leq h \leq 10$. The existence of a relevant average
input connectivity $\langle \kappa \rangle$ between $2$ and $3$ for homogeneous networks in the
context of RBNs was pointed out several years ago by Kauffman \cite{kauffman},
who later suggested that this range could be even larger (up to $\langle \kappa \rangle=5$) if
the Boolean functions are biased toward higher internal homogeneity
\cite{harris}. Our results provide another possible scenario for such a high
value of $\langle \kappa \rangle$. Namely, that heterogeneous distributions together with highly
nonlinear interactions allow for larger $\langle \kappa \rangle$ values. Moreover, contrary to
what has been observed in RBNs, $p_1(h)$ depends only slightly on the system
size, which makes the previous analysis meaningful for larger networks.

As for metabolic networks, the system of differential equations, Eqs.\ 
(\ref{eq1}-\ref{eq2}), represents one of the most basic biochemical reactions,
where substrates and enzymes are involved in a reaction that produces a given
product. In this context, there are several important issues as how fast the
equilibrium is reached, how the concentration of substrates and enzymes
compare, etc. Besides, it is known that in a large number of situations, some
of the enzymes involved show periodic increments in their activity during
division, and these reflect periodic changes in the rate of enzyme synthesis.
This is achieved by regulatory mechanisms that necessarily require some kind of
feedback control as that emerging in our model. Obviously, there is a vast
literature on this kind of process, but the point here is that the real
topological features of the underlying metabolic network \cite{metab} has not
been taken into account, and as far as dynamics is concerned, they should be
incorporated in current models. We are planning to address this issue in the
future.

In summary, we have studied a Michealis-Menten like dynamical model on top of
complex heterogeneous networks with directed links. The free parameters of the
model allowed us the full exploration of the phase diagram of the system's
dynamics under different dynamical and structural conditions. The results
obtained point to a rich behavioral repertoire with stationary, periodic and
chaotic regimes.  A direct comparison with experiments is a tough task,
especially, because experimental results are only now becoming available.
However, we anticipate several features of interest such as that heterogeneous
networks reduces the parameter range in which the existence of chaotic behavior
is likely and that distinct behaviors correlate with connectivity classes.
These suggestions could be tested when more experimental information becomes
available. Our study may help understand biological processes such as gene
expression and reaction kinetics with the tools of network modeling and
nonlinear dynamics. Finally, we would like to point out that further numerical
simulations with other random scale-free networks with lower exponents $\gamma$
(i.e., more heterogeneous nets) \cite{us2}, suggest that the main conclusions
drawn here do not significantly depend on $\gamma$.

\section{Acknowledgments}

We thank F.\ Falo, P.\ J.\ Mart\'{\i}nez and
J. L. Garc\'{\i}a-Palacios for fruitful discussions and
suggestions. Y.\ M.\ thanks S.\ Boccaletti for suggestions. J.\ G-G\
acknowledges financial support of the MECyD through a FPU grant. Y.\
M.\ is supported by a BIFI Research Grant and by MCyT through the
Ram\'on y Cajal program. This work has been partially supported by the
Spanish DGICYT projects BFM2002-01798 and BFM2002-00113.


\begin{thebibliography}{0}
\bibitem{note1} Since the integration is done at fixed time intervals, a
  further numerical check is impossed. Namely, $|g_i(t)-g_i(0)|<
  \frac{dg_i(0)/dt}{2}$ and $|\frac{dg_i(t)}{dt}-\frac{dg_i(0)}{dt}|<
  \frac{d^2g_i(0)/dt^2}{2}$.
\end{thebibliography}
\end{document}